\documentclass[structabstract]{aa}
\usepackage{graphicx}
\usepackage{txfonts}
\usepackage{natbib}

\bibpunct{(}{)}{;}{a}{}{,} 

\begin{document}
\title{S-process in low-mass extremely metal-poor stars}
\author{M. A. Cruz\inst{\ref{inst1}}
\and A. Serenelli\inst{\ref{inst2},\ref{inst1}}
\and A. Weiss\inst{\ref{inst1}}}
\institute{Max-Planck Institut f\"ur Astrophysik,
  Karl-Schwarzschild-Str.     1,     85748,     Garching    bei     M\"unchen,
  Germany\label{inst1} 
\and Instituto de Ciencias del Espacio (CSIC-IEEC), Facultad de
Ci\`encies, Campus UAB, 08193 Bellaterra, Spain\label{inst2} }

\abstract {Extremely  metal-poor (EMP) low-mass stars  experience an ingestion
  of protons into the helium-rich layer during the core He-flash, resulting in
  the production  of neutrons through  the reactions $\mathrm{^{12}C(p,\gamma)
    ^{13}N(\beta)^{13}C(\alpha,n)^{16}O}$.  This  is a potential  site for the
  production of s-process elements in extremely metal-poor stars not occurring
  in  more  metal-rich   counterparts.   Observationally,  the  signatures  of
  s-process elements  in the two most  iron deficient stars  observed to date,
  HE1327-2326 \& HE0107-5240, still await for an explanation.} {We investigate
  the possibility  that low-mass  EMP stars could  be the source  of s-process
  elements observed in extremely iron deficient stars, either as a
  result of  self-enrichment or in a  binary scenario as the  consequence of a
  mass  transfer  episode.}   {We  present  evolutionary  and  post-processing
  s-process  calculations  of  a  $\mathrm{1~M_{\odot}}$  stellar  model  with
  metallicities  $\mathrm{Z=0,\   10^{-8}}$  and  $10^{-7}$.   We  assess  the
  sensitivity of nucleosynthesis results to uncertainties in the input physics
  of  the stellar  models, particularly  regarding the  details  of convective
  mixing during  the core He-flash.}   {Our models provide the  possibility of
  explaining the  C, O, Sr, and Ba  abundance for the star  HE0107-5240 as the
  result of mass-transfer from a low-mass EMP star.  The drawback of our model
  is that if  mass would be transfered before the primary  star enters the AGB
  phase,  nitrogen  would  be  overproduced and  the  $\mathrm{^{12}C/^{13}C}$
  abundance  ratio  would  be  underproduced  in comparison  to  the  observed
  values. } {Our  results show that low-mass EMP stars cannot  be ruled out as
  the companion stars that might  have polluted HE1327-2326 \& HE0107-5240 and
  produced the observed s-process  pattern.  However, more detailed studies of
  the core He-flash  and the proton ingestion episode  are needed to determine
  the robustness of our predictions.}

\maketitle

\section{Introduction}
\label{sec:intro}

Extremely metal-poor (EMP; $\mathrm{[Fe/H]<-3.0}$ - \citet{Beers05}) stars are
an  important key  for understanding  galactic chemical  evolution  during the
early stages of our Galaxy. In addition, abundance patterns of EMP stars might
shed light on individual nucleosynthesis  processes, contrary to those of more
metal-rich   stars,  which   reflect  the   well-mixed  products   of  several
nucleosynthesis  processes in  multiple  generations of  stars. The  abundance
patterns  of  EMP  stars   observed  with  high  resolution  spectroscopy  are
peculiar. A  most conspicuous finding is  that the fraction of  EMP stars that
are carbon enhanced ($\mathrm{[C/Fe]>0.9}$; \citealt{Masseron10}) is $\sim$ 20\%
\citep{Rossi,Suda11}, much larger than the $\sim$ 1\% found in more metal-rich
stars  \citep{Tomkin,Luck}. Also, in  terms of  neutron capture  elements, EMP
stars have  a wide variety  of patterns, with  stars showing no  signatures of
these  elements to  others having  large enhancements  in either  s-process or
r-process elements, or in  both \citep{Beers05}. Moreover, carbon-enhanced EMP
(CEMP) stars are in  their majority also enriched in s-process elements,
suggesting a common origin for carbon and s-process elements in such stars.

Star formation theory can also benefit  from the study of abundance patters of
EMP  stars. The  longstanding argument  that primordial  stars  were extremely
massive      ($\mathrm{M\sim100M_{\odot}}$)      and     isolated      systems
\citep{Abel,Bromm02,Bromm04,Oshea,Yoshida08}   is  being  confronted   by  new
numerical  simulations  of  Population  III  star  formation.  Simulations  by
\citet{Greif} seem to  indicate that Pop. III stars  can be formed in
multiple systems 
with         a         flat         protostellar         mass         function
($\mathrm{M\sim0.1-10M_{\odot}}$).  Moreover, the recent  discovery of  an EMP
star  with  metallicity  $\mathrm{[Fe/H]}=-4.89$  and no  enhancement  in  CNO
elements  \citep{Caffau2011,Caffau2012},  which results  in  the lowest  total
metallicity $\mathrm{Z  \leq 10^{-4}Z_{\odot}}$ observed in a  star, also puts
into question the  claims of a minimum metallicity  required for the formation
of low-mass  stars \citep{Bromm03,Schneider2003}. These  results reinforce the
necessity  of determining  the mass  range of  the stars  responsible  for the
chemical signatures observed  in EMP stars, in order  to constrain the initial
mass function in the early Universe.
   
Several attempts have  been made in order to explain  the peculiarities of EMP
abundance  patterns;   their  origin,  nevertheless,  is   still  an  unsolved
puzzle. One  of the first scenarios  proposed to explain  the overabundance of
carbon and  nitrogen in EMP stars  involved the ingestion of  protons into the
helium convective zone (proton  ingestion episode-PIE) that happens during the
core       He-flash      in       Pop.      III       and       EMP      stars
\citep{Fujimoto90,Schlattl01,Weiss04,Picardi}.  The proton ingestion  into the
helium-rich convective layers during the  core He-flash is a robust phenomenon
in  one dimensional stellar  evolution calculations  of low  metallicity stars
($\mathrm{[Fe/H]<-4.5}$) that subsequently leads to a large surface enrichment
in  CNO-elements \citep{Hollowell,Schlattl01,Campbell08}. However,  the amount
of C, N, and O dredged-up to the surface in the models is usually too large (1
to  3  dex  larger  than  the  observed [C/Fe])  to  match  the  observations,
disfavoring the self-enrichment scenario. Besides, most EMP stars observed are
not evolved past the helium-core flash to have undergone PIE.

The positive correlation between carbon  and barium abundances observed in EMP
stars  indicates  that carbon  and  s-process  elements  might have  a  common
production   site   \citep{Aoki02,Suda04,Suda11}.    Stars  with   metallicity
$\mathrm{[Fe/H] \le  -2.5}$ and in the  mass range $\mathrm{1.2M_{\odot}<M\leq
  3.0M_{\odot}}$ also  experience a  PIE during the  first one or  two thermal
pulses  in the beginning  of the  thermally pulsating  AGB (TP-AGB)  phase. In
analogy to  the occurrence  of the  PIE during the  core He-flash,  during the
development  of a  thermal runaway  in the  He-shell, proton-rich  material is
dragged down into the He-burning  zone and, after the ensuing dredge-up event,
the  stellar envelope  is  strongly  enriched in  CNO  elements (and  possibly
s-process).  An  alternative scenario,  based on the  previous considerations,
was  proposed by \citet{Suda04}.   They argued  that this  is the  most likely
production  environment  for  carbon  and s-process  elements.   The  observed
abundances would be,  then, the result of mass transfer  from an AGB companion
(now  a white  dwarf).   Radial-velocity  monitoring has  shown  that a  large
fraction  of  the CEMP-s  (the  subset of  CEMP  stars  enriched in  s-process
elements) presents variable radial velocities, strengthening the argument of a
binary scenario for the formation of CEMP-s stars \citep{Lucatello05}.

Assuming that carbon and s-process enhancements in CEMP-s stars are the result
of mass-transfer from  a primary star, the question  remains whether the donor
stars  are  low-  or   intermediate-mass  stars.   \citet{Suda04}  claim  that
intermediate-mass stars are the  polluting stars, responsible for the observed
enhancements    in     carbon    and    s-process     elements    in    CEMP-s
stars. \citet{Campbell10} have calculated  models of low-mass EMP stars during
the core He-flash  and found that the  PIE leads to the production  of a large
amount of neutrons  ($\mathrm{n_n>10^{14} cm^{-3}}$) lasting for $\mathrm{\sim
  0.2}$ years.  Consequently, they have found production of s-process elements
during the core He-flash of  such star.  On the other hand, \citet{Fujimoto00}
and \citet{Suda10} claim  that no s-process signature would  be present in the
photosphere  of  low-mass  stars  that   suffered  the  PIE  during  the  core
He-flash. A possible  explanation for the difference between  these studies is
the    shorter     duration    of    the     PIE    in    the     models    of
\citet{Suda10}. One-dimensional  stellar evolution calculations of  the PIE by
all  authors show  that,  after enough  protons  have been  ingested into  the
He-burning  region,  the convective  He-burning  region  splits  into a  lower
He-burning convective zone and an  upper H-burning region (later to merge with
the stellar  envelope).  However, conditions  in the H-burning layers  are not
favourable for s-process  to happen. Because the convective  zone in models by
\citet{Suda10}  splits before s-process  can take  place no  s-process surface
enrichment is  observed.  The  current results in  the literature for  the PIE
during the  core He-flash are inconclusive  with respect to  the part low-mass
EMP stars  play in the s-process  production in the  early Universe.  Although
the  PIE is  a well  established phenomenon  in current  1D  stellar evolution
modeling, different studies  do not agree about its  main properties, and this
is reflected in the production of s-process elements.

The aim of the present work is to explore the s-process production by low-mass
EMP stars and their role as  a source of the s-process enhancement observed in
EMP  stars.   We  investigate  the  main properties  affecting  the  s-process
production in these stars.  We present evolutionary and s-process calculations
for the  core He-flash of $\mathrm{1~M_{\odot}}$ stellar  models. Our approach
differs from  the \citet{Campbell10}  in that we  use a more  complete network
including   a  neutron   sink   in  the   evolutionary   calculations  and   a
post-processing  code   for  {\em  a  posteriori}   calculation  of  s-process
nucleosynthesis  instead of solving  the full  network (including  the neutron
releasing reactions)  in a post-processing code.   Also, we have  been able to
follow  the evolution  of our  models through  the entire  PIE  and subsequent
dredge-up phase. This has allowed us to directly obtain the surface enrichment
in  the models.   On the  contrary, \citet{Campbell10}  estimated  the surface
enrichment  from  previous  simulations  (using  a  smaller  network)  due  to
numerical problems in following the subsequent dredge-up. We hope to close the
gaps left  by \citet{Campbell10} by presenting models  including the evolution
of the star during the dredge-up episode, preventing in this way the necessity
of extrapolating the resulting envelope abundances for all elements.

The   paper  is  organized   as  follows:   In  Sects.    \ref{sec:codes}  and
\ref{sec:models}  we describe  the input  physics of  the codes  used  and the
models calculated,  respectively. We discuss  the main characteristics  of the
proton ingestion episode in Sect. \ref{sec:pie} and the subsequent neutron and
associated   s-process  production   during   the  core   He-flash  in   Sect.
\ref{sec:sprocess}.  We   compare  our   results  with  the   observations  in
Sect. \ref{sec:discussion}.

\section{Stellar evolution and nucleosynthesis code}
\label{sec:codes}

In the calculations presented here we have used the GARching STellar Evolution
Code, GARSTEC, as described in  \citet{weiss}. Convection is modeled using the
Mixing Length Theory  (MLT) and the mixing length  parameter $\mathrm{\alpha =
  1.75}$, similar to the one  obtained from standard solar model calculations.
Mass-loss    is    included    using   \citet{reimers}    prescription    with
$\mathrm{\eta=0.4}$. OPAL equation of state \citep{Rogers} was used. Radiative
opacity tables are  also from OPAL \citet{Iglesias}. At  low temperatures they
are complemented by  those from \citet{Ferguson} and extended  sets to account
for carbon-rich compositions (see  \citet{Weiss09} for details).  We have used
the results by \citet {Itoh} for electron conduction opacities.

For   the  present   work,   the   nuclear  network   in   GARSTEC  has   been
extended. Changes are described in detailed in the next section.

\subsection{Code implementation}

The  version of  GARSTEC from  \citet{weiss} follows  the evolution  of stable
isotopes involved  in the  p-p chain,the CNO  cycles, and the  standard helium
burning reactions. For  this work the network in GARSTEC  has been extended to
include all relevant nucleosynthesis  processes for intermediate mass elements
($\mathrm{A<30}$)  during H-burning.  This  is important  because it  has been
suggested that in metal-free primordial  stars elements in this mass range can
be produced  during H-burning (e.g. by NeNa  and MgAl chains) and  then act as
seeds  for the s-process  \citep{Goriely01}. Also,  reactions that  govern the
production of  neutrons during  the He-burning have  been incorporated  in the
network. In total, the network  comprises 34 isotopes linked by 120 reactions,
including proton, alpha, and neutron captures, and beta decays.

For charged-particle reactions the rates are from the NACRE database
\citep{Angulo} and from the JINA REACLIB library \citep{Cyburt},
except for: 
\begin{itemize}
\item $\mathrm{^{14}N(p,\gamma)^{15}O}$ - \citet{Adelberger};
\item  $\mathrm{^{17}O(p,\gamma)^{18}F}$ and $\mathrm{^{17}O(p,\alpha)^{14}N}$
  - \citet{Moazen}; 
\item $\mathrm{^{22}Ne(p,\gamma)^{23}Na}$ - \citet{Hale1};
\item $\mathrm{^{23}Na(p,\gamma)^{24}Mg}$ and $\mathrm{^{23}Na(p,\alpha)^{20}Ne}$ -
  \citet{Hale2}; 
\item $\mathrm{^{13}C(\alpha,n)^{16}O}$ - \citet{Kubono};
\item $\mathrm{^{22}Ne(\alpha,n)^{25}Mg}$ - \citet{Jaeger};
\item $\mathrm{^{12}C(\alpha,\gamma)^{16}O}$ - \citet{Kunz}. 
\end{itemize}

Experimental  rates   for  neutron  capture  reactions  are   taken  from  the
recommended         values         given         in        the         KADONIS
database\footnote{http://www.kadonis.org/} \citep{Dillmann}.  For isotopes for
which  experimental data  is  not available,  rates  are taken  from the  JINA
REACLIB.\footnote{https://groups.nscl.msu.edu/jina/reaclib/db/}

In  order  to  accurately  compute   the  abundance  of  neutrons  during  the
evolutionary calculations so that it can be used for post-processing 
calculations, the effect of neutron captures on all isotopes not 
included in the network has to be accounted for by using a neutron sink, which we
call $\mathrm{^{30}AA}$, as described in \citet{Jorissen}. The number fraction
and the neutron capture cross-section of the sink element are given by
\begin{equation}
Y_{AA} = \sum_{i={\rm ^{30}Si}}^{{\rm ^{211}Po}} Y_{i}
 \label{numfrac}
\end{equation}
\begin{equation}
\sigma_{(AA,n)}(\tau)= \sum_{i={\rm ^{30}Si}}^{{\rm ^{211}Po}} \frac{Y_{i}(\tau)\sigma_{(i,n)}}
{Y_{AA}},
 \label{numfrac2}
\end{equation}
respectively, where  the summation  extends over  all  the relevant
isotopes  not  included in  the  evolutionary  calculations. The  sink
  cross-section   ($\mathrm{\sigma_{AA,n}(\tau)}$)  depends  on   the  neutron
  capture path which is represented  in this equation by the neutron exposure.
  The neutron  exposure $\mathrm{\tau}$ is  given by $\mathrm{\tau(t)=\int_0^t
    n_{n}(t')v_{T}dt'}$, where $\mathrm{n_{n}}$  is the neutron number density
  and $\mathrm{v_{T}}$ is the thermal velocity of neutrons at temperature T.

One of the difficulties in the neutron sink approach is the calculation of the
sink  cross-section  $\mathrm{\sigma_{(AA,n)}}(\tau)$,   which  has  to  be  known
beforehand  in order  to  get accurate  estimates  of the  number of  neutrons
captured   by  heavy   elements  in   the  evolutionary   calculations.   This
cross-section,  however, depends on  the distribution  of heavy  elements, and
therefore on  the neutron capture path. Thus,  detailed s-process calculations
are necessary to estimate values for $\mathrm{\sigma_{(AA,n)}}(\tau)$.

In AGB  stars with metallicity $\mathrm{[Fe/H]>-2.5}$, s-process  happens in a
radiative  environment  and  the  typical  neutron exposure  is  smaller  than
$\mathrm{1.0}$  $\mathrm{mb^{-1}}$. \citet{Herwig03} have  shown that,
  in  the solar  metallicity  case, the  use of  a  fixed value  for the  sink
  cross-section  in the  calculations with  the sink  treatment leads  to time
  evolution of  the neutron  exposure very similar  to that obtained  with the
  calculations  including  all  isotopes  up  to Pb.  In  their  calculations,
  they     found     that    a     sink     cross-section
  $\mathrm{\sigma_{(AA,n)}}$  of  $120\,   \mathrm{mb}$  could  reproduce  the
  neutron  exposure  evolution from  the  full  s-process calculations  within
  10\%.

Conditions in EMP stars during the PIE are, however, quite different from
  those 
during the  canonical thermal  pulses in  AGB stars. In  the PIE  neutrons are
produced and consumed  in convective regions.  The large  ingestion of protons
that  occurs  in EMP  stellar  models, either  during  the  core He-flash  for
low-mass  stars or  early in  the AGB  for intermediate-mass
stars leads  to the  production of a  significant amount  of $\mathrm{^{13}C}$
through  the reaction $\mathrm{^{12}C(p,\gamma)  ^{13}N(\beta^{-})^{13}C}$ and
might  result in neutron  exposures much  larger than  those achieved  in more
metal-rich stars.  Therefore, in  order to compute $\sigma_{\rm (AA,n)}(\tau)$
as a 
function of  the neutron irradiation  $\tau$ (eq.\ref{numfrac2}), and  to test
the  accuracy of  the sink  approach during  the PIE  in reproducing  the time
evolution of the neutron exposure, we have performed a number
of parametrized s-process calculations that we describe below.

For   the  parametrized   s-process  calculations   we   take  constant
  temperature and density values typically found during the
  PIE. We have performed a number of calculations covering the ranges
  $\mathrm{T=1.0-2.5          \times10^          {8}\         K}$          and
  $\mathrm{\rho=100-5000\ g\ cm^{-3}}$.  First, in order to compute $\sigma_{\rm
    (AA,n)}(\tau)$ we perform the network integrations using a full s-process
  network with 613 isotopes from  neutrons to $^{211}$Po and including proton,
  neutron and alpha captures for all  isotopes with ${\rm A < 30}$ and neutron
  captures for those with ${\rm A \geq 30}$. In all cases,  abundances of isotopes
  with ${\rm A<30}$ are taken from a stellar model at the moment where the PIE
  takes  place.  For  isotopes with  ${\rm A\geq  30}$, we  take  scaled solar
  abundances \citep{grevesse:1998}. The only  exception is $^{13}$C, for which
  we took  two different  mass fraction values:  $10^{-2}$ and  $10^{-4}$. The
  outcome of these calculations is the  time evolution of the abundances of all
  isotopes and, therefore, they allow a direct evaluation of $\sigma_{\rm (AA,n)}$
  as a  function of time  (or neutron exposure; equation~\ref{numfrac2}.  As a
  second step,  we repeat the calculations  but using now  the smaller network
  employed in GARSTEC  for the evolutionary models and  using, for the neutron
  sink, the cross section derived in the first step. The comparison of results
  of the neutron exposure in the first and second steps allows us to check the
  accuracy with which  this quantity can be reproduced  in simplified networks
  that use a neutron sink element.

\begin{figure*}
  \resizebox{\hsize}{!}{\includegraphics{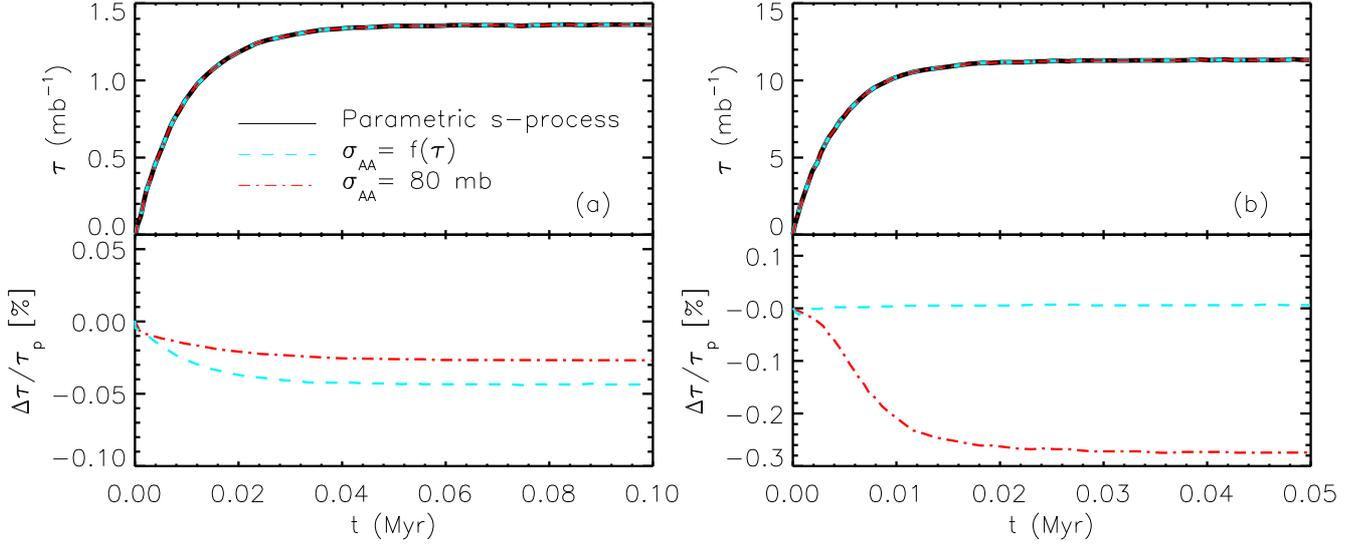}}
  \caption{Top panels: time evolution of the neutron exposure for the parametrized
    s-process  calculation  and  the  neutron  sink  approach  with  different
    cross-section values for a composition of heavy elements ($\mathrm{A>30}$)
    corresponding   to   a   solar    abundance   pattern   scaled   down   to
    $\mathrm{Z=10^{-8}}$.   The   abundance    values   for   light   elements
    ($\mathrm{A<30}$) were taken from  stellar models calculated using GARSTEC
    except  for $\mathrm{^{13}C}$  for  which we  take  two different  values.
    Panel (a) shows results for a $\mathrm{^{13}C}$ mass-fraction of $10^{-4}$
    and panel (b) for a $\mathrm{^{13}C}$ mass-fraction of $10^{-2}$.
      In both
    cases, lines depicting results for the full network (black line) 
    and a sink element (cyan and red lines) overlap. To
    enhance  differences we  show,  for each  value  of the  $\mathrm{^{13}C}$
    abundance, the  relative difference in percentage of  the neutron exposure
    with  respect   to the  neutron  exposure   from  the  parametric
      calculations ($\tau_{\rm p}$) overlap in both 
    plots. \label{figure:1}}
\end{figure*}

Results are illustrated in
  Figure~\ref{figure:1} where we show, for the case $\mathrm{T=10^{8}~K}$ and
$\mathrm{\rho=   500\  g\   cm^{-3}}$  and   the  two   choices   of  $^{13}$C
abundance, the time evolution of the neutron exposure. The neutron exposure 
derived using the full
s-process network is shown in black solid line. Results using the small
network  plus  the  neutron sink  with  the  capture  cross section  given  by
Equation~\ref{numfrac2} are shown in dashed  cyan line. The bottom panels show
relative differences.  Clearly, the sink approach performs very well and 
reproduces the  neutron exposure  of the full  network calculation to  a 0.1\%
accuracy.  Additionally, we performed a test to check the sensitivity of
neutron  exposure   predictions  to   the  specific  choice   of  $\sigma_{\rm
  (AA,n)}$. For this exercise, we adopt the very simplistic approach of 
fixing $\sigma_{\rm (AA,n)}=80\mathrm{mb}$, i.e. to a constant value.
Results  are shown  in  dash-dotted  red
line. Even in this case the neutron exposure is accurately reproduced to about
0.3\%. These results are  representative of all parametrized calculations with
different T and $\rho$ values.

These results are much better than  those found in the less extreme conditions
present in AGB stars, as mentioned in previous paragraphs.  The reason is 
quite simple. When protons are ingested into the helium-burning region during the
PIE,  light  metals  have   already  been  produced.   These  primary  metals,
predominantly $^{12}$C,  amount to a  mass fraction of  about 0.05 and  they are
about  $10^8$ times more  abundant than  metals with  A$\geq$30. The  PIE only
takes place in extremely metal-poor  models.  Taking a typical average neutron
cross-section  for  the light  and  heavy  elements  of 0.01~mb  and  1000~mb,
respectively,  the  average  contribution  to  the neutron  capture  by  light
elements  is much  larger  than by  heavy  elements in  such environment.   We
estimate    $\mathrm{\sigma_{light}X_{light}/   \sigma_{heavy}X_{heavy}   \sim
  1500}$.  Almost  all neutrons  are captured by  elements with A$<30$  in the
environment associated  with the PIE,  and these capture reactions  are always
treated in detail in our models.  Therefore, the value of $\sigma_{\rm(AA,n)}$
used during a  PIE is of little consequence for  predicting  a neutron
  exposure consistent  with that derived  from full network  calculations.
 
In  this  work,  all   post-processing  calculations  are  based  upon
  evolutionary 
calculations where the neutron capture cross section of the sink is given as a
function of the neutron exposure, i.e. as derived from Equation~\ref{numfrac2}.
For each parametrized calculation, a fitting formula for $\sigma_{\rm (AA,n)}$
as  a function  of $\tau$  is derived.  These formulae  are then  used  in the
evolutionary calculations to compute the neutron capture cross section of the
sink element.  In any  case, as  just discussed, the  detailed choice  of this
quantity is of very little consequence for the s-process calculations.

\subsection{Post-processing program}

The  post-processing  code uses  the  model  structure (temperature,  density,
neutron abundance, and convection  velocity) from evolutionary calculations as
input for the  s-process calculations.  The calculations are  done between two
structural models and  the evolutionary timestep between them  is divided into
smaller burning and mixing steps.  All structural quantities are kept constant
between the two stellar models.  Within each substep, we first compute changes
in chemical  composition due  to nuclear burning  and then mix  the convective
(and overshooting) regions.

We adopted the time-dependent mixing scheme described by
\citet{Chieffi}:
\begin{equation}
\mathrm{X_{i}^{k}} = 
\mathrm{^{0}X_{i}^{k}+\frac{1}{M_{mixed}}\sum_{j= mixed}(^{0}X_{j}^{k} -} 
\mathrm{ ^{0}X_{i}^{k})f_{ij}\Delta M_{j}},
 \label{mixsch}
\end{equation}
\noindent where the sum extends  over the mixed region (including overshooting
regions if present), $\mathrm{\Delta M_{j}}$ is the mass of the shell {\it j},
$\mathrm{^{0}X_{i}^{k}}$  and $\mathrm{^{0}X_{j}^{k}}$ are,  respectively, the
abundances of isotope {\it k} in the shells {\it i} and {\it j} before mixing,
$\mathrm{M_{mixed}}$   is  the   mass  of   the  entire   mixed   region,  and
$\mathrm{f_{ij}}$ is a damping factor given by
\begin{equation}
\mathrm{f_{ij}=min(\frac{\Delta t}{\tau_{ij}},1)}
\label{damp}
\end{equation}
that accounts  for partial mixing between zones  {\it i} and {\it  j} when the
mixing  timescale  between  them,  $\mathrm{\tau_{ij}}$, is  longer  than  the
timestep   used    in   the   calculations,    $\mathrm{\Delta   t}$.    Here,
$\mathrm{\tau_{ij}}$  is  assumed to  be  the  convective turn-over  timescale
between  grid points  {\it  i} and  {\it  j} and  the  convective velocity  is
determined using MLT. With this scheme we account for partial mixing occurring
when the  convective turn-over  timescale between two  shells in the  model is
longer than the  mixing timestep. This scheme  is very easily implemented
  and,  importantly,   much  faster  than  the  diffusive   approach  used  in
  GARSTEC. We  have checked,  with a test  version of the  postprocessing code
  where the  mixing is  treated by solving  the diffusion equations,  that for
  sufficiently small timesteps, both schemes agree very well.

Weak interaction rates (electron captures and $\beta$ decays) are taken mainly
from \citet{Taka} and  interpolated as a function of  temperature and electron
density.  At  temperatures lower  than $10^{7}$ K  we have assumed  a constant
value equal to the laboratory rate.  Sources for the neutron capture rates are
the same used  for the network in GARSTEC, either the  JINA REACLIB library or
calculated using the newest KADONIS results. Proton and alpha captures are not
included in  the network, since  in the post-processing only  isotopes heavier
than $\mathrm{^{29}Si}$ are  included and the relevant reactions  for them are
the neutron captures and beta decays.

The network includes 580  isotopes (starting from $\mathrm{^{30}Si}$) and more
than 1000  reactions. Branching points  involving the most  well-known isomers
are  included  and the  network  terminates with  the  $\alpha$  decays of  Po
isotopes.  All isotopes with decay lifetimes comparable to or longer than
  the lifetime against neutron captures have been included in the network.

\section{The Models}
\label{sec:models}

We  performed  evolutionary  calculations  of  $\mathrm{1~M_{\odot}}$  stellar
models from  the zero-age main  sequence to the  dredge-up after the  PIE. The
composition is solar-scaled with metallicities $Z=0$ and $\mathrm{Z=10^{-8}}$.
Our ``standard'' models (M1 and M2  - see Table \ref{table:1}) do not include
  overshooting or metal diffusion. 

Mixing in convective regions is  modeled as a diffusive process. The diffusion
coefficient  in convective  regions is  taken as  $\mathrm{D_{c}=1/3  v \times
  \ell}$,  where   v  is  the   convective  velocity  derived  from   MLT  and
$\mathrm{\ell}$ is  the mixing length. It  has been suggested  that during the
PIE, the  reactive nature of the  hydrogen material adds some  buoyancy to the
sinking  elements and  this  results in  a  reduced efficiency  of the  mixing
\citep{Herwig11}. We have mimicked this situation by using a reduced diffusion
coefficient  by   dividing  the  above  relation   by  10  in   a  model  with
$\mathrm{Z=10^{-8}}$ (model M3 - Table \ref{table:1}).

We   have  also   computed   additional   models  with   metallicity
$\mathrm{Z=10^{-8}}$ including overshooting.  Overshooting has been modeled as
an   exponentially   decaying   diffusive   coefficient,   as   described   in
\citet{Freytag}, and has been included  in all convective boundaries. In order
to test  the influence and associated uncertainties  of this poorly-understood
phenomenon, two values of the free parameter $f$ that determines the extension
of the  overshooting regions have been  chosen: $f=0.016$ and  0.07 (models M4
and M5  respectively -  Table \ref{table:1}).  The  first value is  a standard
choice  for this  prescription  of overshooting  because  it gives  comparable
results in  the main sequence  as the canonical  choice of 0.2  pressure scale
heights \citep{Magic10},  a value  known to reproduce  different observational
constraints such as the width of the main sequence observed in different
open  clusters.   The higher  value  is intended  to  represent  a case  where
overshooting  is  much  larger  than  in convective  cores  and  deep  stellar
envelopes, and closer to what was originally found by \citet{Freytag} for thin
convective  envelopes   of  A-type   stars.   Incidentally,  this   amount  of
overshooting gives a lithium depletion in solar models
\citep{Weiss99} consistent with observations.

Finally, we have computed an additional model, Z7, with the same physical
  input as model M2 but  an initial metallicity $\mathrm{Z=10^{-7}}$, i.e. one
  order of  magnitude larger.  This allows  us to consider  the effect  of the
  initial  metallicity on  the  production of  elements  as well  as a  closer
  comparison to HE0107-5240.

\section{Main Characteristics of the Proton Ingestion Episode}
\label{sec:pie}

In this and following sections we  will use model M2 (See Table \ref{table:1})
to illustrate the characteristics of the PIE, unless stated otherwise.

\begin{figure}
  \resizebox{\hsize}{!}{\includegraphics{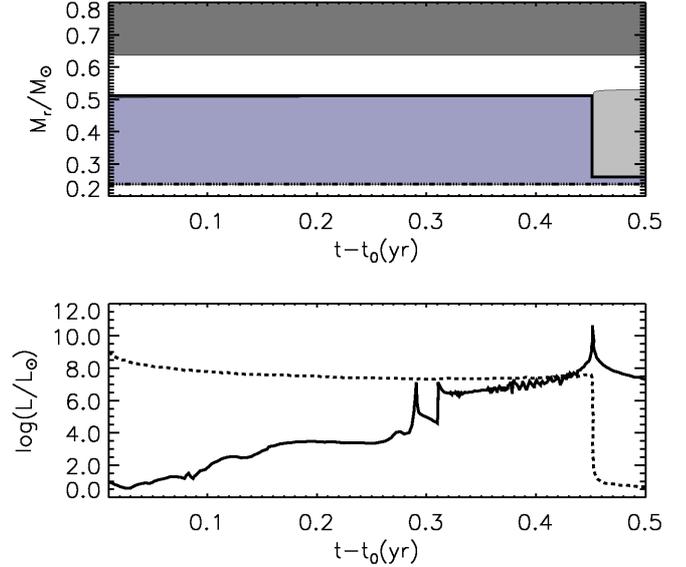}}
  \caption{Upper Panel: Time evolution of convective zones in model M2
      $\mathrm{Z=10^{-8}}$ during the PIE. Dark grey area: convective
      envelope, purple area between 0.24 and 0.51~M$_\odot$: HeCZ, light grey
      area: HCZ.  Thick
      solid line represents the position where hydrogen mass fraction is 0.001;
      dot-dashed  line represents  the position  of  maximum energy
      release due to He-burning, respectively. The splitting of
      the HeCZ  around $t-t_0= 0.451~\hbox{yr}$ is evident  with the formation
      of  the  detached HCZ.   Bottom Panel:  Evolution  of
      H-burning  (solid  line)  and  He-burning  (dashed  line)  luminosities.
      $t=t_0$ corresponds to the time of maximum He-burning luminosity.}
  \label{figure:2}
\end{figure}

Low-mass stars  ignite He-burning under degenerate conditions,  leading to the
well known  core He-flash. Due to  stronger neutrino cooling  in the innermost
core the  ignition point is  off-center, as shown  by the position  of maximum
energy release by He-burning $\mathrm{M^{He}_{max}=0.237\ M_{\odot}}$
(Fig.~\ref{figure:2}). Ignition of He-burning  results in the formation of the
so-called helium convective  zone (HeCZ) the outer boundary  of which advances
in mass during  the development of the He-flash.   In Figure~\ref{figure:2} we
show the evolution of the HeCZ (purple region extending from
$M_r/M_\odot=0.237$  up  to  0.510)  during  the time  following  the  maximum
helium-burning luminosity  up to the  moment where the  PIE starts.  The
combination of  an off-center  He-burning ignition and  a low  entropy barrier
between the He- and H-burning shells in the very metal-poor environment allows
the HeCZ  to reach hydrogen-rich  layers \citep{Hollowell} and giving  rise to
the PIE.

\begin{figure}
  \resizebox{\hsize}{!}{\includegraphics{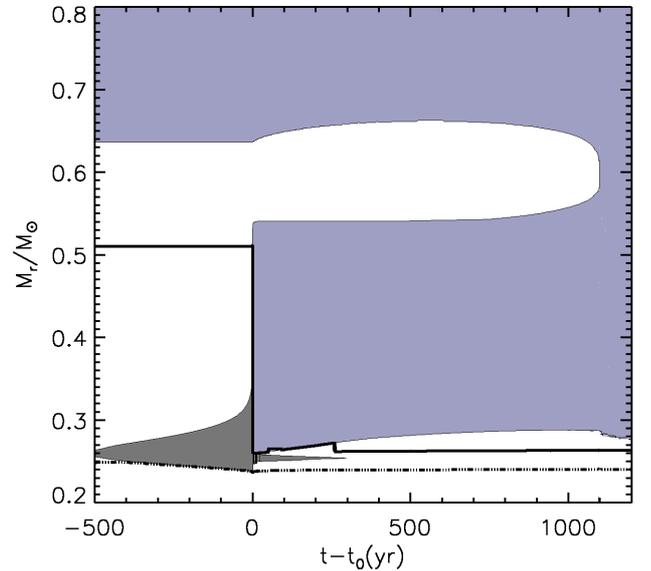}}
  \caption{Long-term evolution  of the 1~M$_\odot$,  Z=10$^{-8}$ stellar model
    during the development  of the core He-flash, the  PIE, and the subsequent
    dredge-up  event. $t=t_0$ corresponds  to the  time of  maximum He-burning
    luminosity. As in Figure~\ref{figure:2},  thick solid and dot-dashed lines
    represent  the position  of maximum  energy release  due to  H-burning and
    He-burning,  respectively. Merging  of the  HCZ and  the  stellar envelope
    occurs around 1100~yr after the PIE takes place.}
  \label{figure:3}
\end{figure}

The HeCZ  reaches the radiatively  stratified H-rich layers on  top (initially
represented  by  white  region   above  $M_r/M_\odot=0.510$)  soon  after  the
He-burning luminosity reaches  its maximum value. Protons are  then mixed down
into the  HeCZ and  are captured by  the abundant  $\mathrm{^{12}C}$, creating
$\mathrm{^{13}C}$. The H-burning luminosity increases while the HeCZ continues
expanding (in mass).  Note the penetration  of the HeCZ into the H-rich region
occurs in a narrow region of only  a $\sim 10^{-3}\, {\rm M_\odot}$  and
it is  barely visible around 0.510~M$_\odot$  in Figure~\ref{figure:2}. Around
 0.451 years  after the  He-luminosity reached  its maximum  a secondary
flash happens, the H-flash, with  a peak H-burning luminosity comparable
to  the He-burning  luminosity ($\mathrm{\log{L^{He}_{max}/L_{\odot}}} =
10.70$, $\mathrm{\log{L^{H}_{max}/L_{\odot}}} = 10.71$).  The burning
timescale  for  hydrogen   through  the  $\mathrm{^{12}C(p,\gamma)}$  reaction
becomes shorter as  protons travel downwards in the HeCZ  until a point inside
the HeCZ  is reached where the  burning and mixing  timescales are comparable.
At this position ($\mathrm{M^{H}_{max} =0.259\,M_{\odot}}$) the energy release by
H-burning is maximum and temperature  rises quickly, resulting in an inversion
in  the temperature  profile that  eventually leads  to the  splitting  of the
convective zone.  This prevents the  penetration of protons further  down into
the HeCZ.

\begin{figure}
  \resizebox{\hsize}{!}{\includegraphics[angle=90]{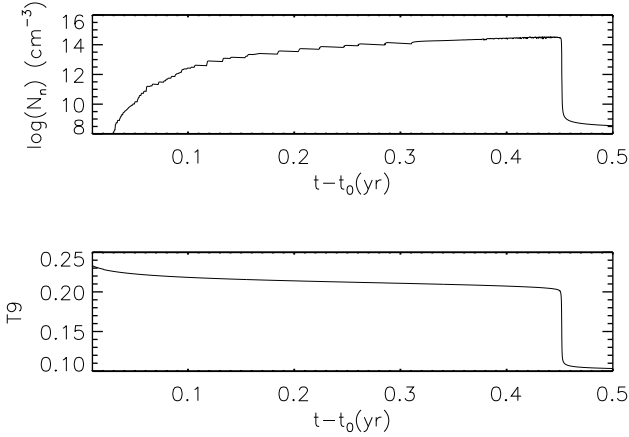}}
  \caption{Upper  Panel: Time  evolution  of maximum  neutron  density in  our
    $\mathrm{Z=10^{-8}}$  model (M2)  at M=0.237  $\mathrm{M_{\odot}}$. Bottom
    Panel: Evolution of  the temperature near the bottom of  the HeCZ. In both
    cases  the sudden  change at  $t-t_0\approx 0.451~\hbox{yr}$  reflects the
    splitting of the HeCZ.}
  \label{figure:4}
\end{figure}

The  splitting results  in two  convective  zone: a  small HeCZ  and an  upper
convective zone  which is referred to  as hydrogen convective  zone (HCZ). The
outer  boundary  of   the  HCZ  continues  to  advance   in  mass.   In  about
$\mathrm{1100}$ years,  the convective envelope  deepens and products
of burning 
during  and after  the PIE  are  dredged-up to  the surface.  The envelope  is
enriched mainly  in C, N,  and O. Similarly  to results by other  authors, the
amount of  C and  N brought up  to the  surface is too  large compared  to the
abundances observed in EMP stars  (values in Table \ref{table:2} are more than
3  dex  larger than  the  average  CEMP's  carbon abundance).   Enrichment  in
s-process depends strongly on the  nucleosynthesis before the splitting of the
HeCZ and will  be discussed in the following  section. The long-term evolution
before and after  the PIE is illustrated in  Figure~\ref{figure:3}, where time
has been set to zero at the moment when $\mathrm{L^{He}}$ reaches its maximum.

The  time between  maximum He-burning  luminosity  and the  onset of  hydrogen
mixing ($\mathrm{\Delta t_{mix}}$) in  our calculations is usually larger than
the  interval found in  other studies  \citep{Hollowell,Suda07,Picardi}.  This
might partially  be the result  of the larger  mass between the bottom  of the
HeCZ and the location of base of the H-rich layer in our models. Our predicted
$\mathrm{\Delta t_{mix}}$ varies from  $\mathrm{10^{-2}}$ to 1 year (depending
on the metallicity and the convection efficiency assumed in the models), while
in   other   studies   this    value   ranges   from   $\mathrm{10^{-3}}$   to
$\mathrm{10^{-2}}$ years. The main properties of the core helium flash and the
PIE are summarized  in Table \ref {table:1}. An important  property of the PIE
for the  s-process is the  time between the  onset of hydrogen mixing  and the
splitting  of  the  convective  zone,  here  referred  to  as  $\mathrm{\Delta
  t_{PIE}}$.    After  the   splitting   of  the   HeCZ,   the  abundance   of
$\mathrm{^{14}N}$, the  main neutron  poison in this  environment, in  the HCZ
rapidly builds  up. As the $\mathrm{^{13}C}$ to  $\mathrm{^{14}N}$ ratio drops
below      unity     neutrons      are      predominantly     captured      by
$\mathrm{^{14}N(n,p)^{14}C}$  and the  s-  process is  effectively shut  down.
Therefore,   $\mathrm{\Delta   t_{PIE}}$    determines   the   timescale   for
s-processing. In  our calculations $\mathrm{\Delta  t_{PIE}}$ is in  the range
$\sim$0.1 to  1 yr (Table~\ref{table:1}),  about an order of  magnitude longer
than  in  some  other   studies  (e.g.   \citealt{Hollowell}).   In  contrast,
\citet{Campbell10} presented  a model with  metallicity $\mathrm{[Fe/H]=-6.5}$
in which $\mathrm{\Delta t_{PIE}\sim 20}$ yrs, about a factor of 50 
larger than our  findings.  According to our  calculations, differences in
  $\mathrm{\Delta t_{PIE}}$  imply differences in the  production of s-process
  elements. Therefore, the large variations found in $\mathrm{\Delta t_{PIE}}$
  just described, which span up to three orders of magnitude, are probably the
  reason why results for  nucleosynthesis of s-process elements from different
  authors range from  no production at all, see  for example \cite{Suda10}, to
  extremely  large production  factors like  in \cite{Campbell10},  as  it was
  described in Section~\ref{sec:intro}.

We note  here that splitting of  the HeCZ is ubiquitous  in all 1-dimensional
simulations   of    the   PIE   during   the   core    helium   flash   \citep
{Hollowell,Schlattl01,Campbell08}. Splitting of the HeCZ (shell) is also found
in  PIE   resulting  from   very  late  thermal   pulses  in   post-AGB  stars
\citep{herwig01,althaus05} as  well as in  simulations of the early  phases of
the TP-AGB in very metal-poor stars \citep {Fujimoto00,sere06,Cristallo09}. On
the other  hand, recent hydrodynamic  3-dimensional simulations of  the TP-AGB
phase for a low-mass star \citep{Stan11} have not shown splitting of the HeCZ,
although the  authors argued this might  be an artifact of  the low resolution
used in the calculations.

\begin{table}
\caption{Main properties of Proton Ingestion Episode}    
\label{table:1}     
\centering                               
\begin{tabular}{c c c c c c c c}       
\hline\hline Model & $\mathrm{\log{L^{He}_{max}}}$\tablefootmark{a} & 
$\mathrm{\log{L^{H}_{max}}}$\tablefootmark{b} & 
$\mathrm{M^{He}_{max}}$\tablefootmark{c} &
$\mathrm{M^{He}_{c}}$\tablefootmark{d} & 
$\mathrm{\Delta 
  t_{mix}}$\tablefootmark{e}   &   $\mathrm{\Delta  t_{PIE}}$\tablefootmark{f}
\\  & $[{\rm  L_\odot}]$  & $[{\rm  L_\odot}]$  & $[{\rm  M_\odot}]$ &  $[{\rm
    M_\odot}]$ & {\rm [yr]} & 
{\rm [yr]} 
\\ \hline M1\tablefootmark{g} & 9.972 & 10.48  & 0.171 & 0.475 & 1.400 &
1.356
\\  M2\tablefootmark{\ }  & 10.70  & 10.71  & 0.237  & 0.508  & 0.020  &
0.432
\\  M3\tablefootmark{h} &  10.69  & 9.736  & 0.236  &  0.508 &  0.027 &
0.482
\\  M4\tablefootmark{i}  &   10.70  &  11.01  &  0.237  &   0.509  &
0.018  &
0.192\\  M5\tablefootmark{j} &  10.69 &  11.03 &  0.237 &  0.509  & $7\times
10^{-3}$ & 0.043 \\ 
 Z7\tablefootmark{k} & 11.10 & 10.37 & 0.267 & 0.529 & 0.200 & 2.204 \\
\hline 
\end{tabular}
\tablefoot{
\tablefoottext{a}{Logarithm of the maximum He-burning.}
\tablefoottext{b}{Logarithm of the maximum H-burning.}
\tablefoottext{c}{Position of maximum energy released by He-burning.}
\tablefoottext{d}{Degenerate core mass.}
\tablefoottext{e}{Time between maximum He-burning luminosity and the
  onset of hydrogen mixing.}
\tablefoottext{f}{Time between maximum He-burning luminosity and the
  splitting of the convective zone.} 
\tablefoottext{g}{Model  with   Z=0.  All   other  models,  except   Z7,  have
  Z=$\mathrm{10^{-8}}$.} 
\tablefoottext{h}{Model  with convective  efficiency  reduced by  an order  of
  magnitude.} 
\tablefoottext{i}{Model including overshooting  with f=0.016.}
\tablefoottext{j}{Model including overshooting  with f=0.07.}
\tablefoottext{k}{Model with initial $Z=10^{-7}$.}
}
\end{table}

\section{Neutron Production and the s-process}
\label{sec:sprocess}

\begin{figure}
  \resizebox{\hsize}{!}{\includegraphics{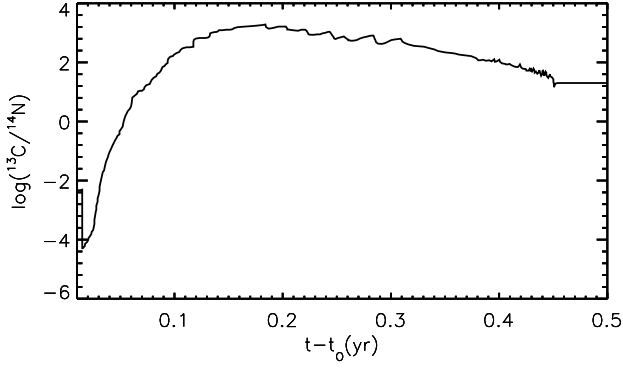}}
  \caption{$\mathrm{^{13}C/^{14}N}$  ratio  near   the  bottom  of  the  HeCZ.
    The   value    of   $\mathrm{t_{0}}$   is   the    same   as   in
    Fig.~\ref{figure:2} and \ref{figure:4}.
  \label{figure:5}}
\end{figure}

During   the  PIE,  $\mathrm{^{13}C}$   is  formed   by  proton   captures  on
$\mathrm{^{12}C}$  and mixed  throughout the  entire HeCZ.   The  abundance of
$\mathrm{^{13}C}$  in  the convective  zone  increases  by  several orders  of
magnitude         during        $\mathrm{\Delta         t_{PIE}}$        (from
$\mathrm{X_{13C}\sim10^{-10}}$     to     $\mathrm{X_{13C}\sim10^{-3}}$     in
mass-fraction).

Neutrons are produced  by the reaction $\mathrm{^{13}C(\alpha,n)^{16}O}$, near
the bottom  of the  convective zone, reaching  maximum neutron  density values
larger than $\mathrm{10^{14} cm^{-3}}$. $\mathrm{^{12}C}$ is the most abundant
isotope, besides $\mathrm{^{4}He}$,  in the HeCZ and thus  neutrons are mainly
captured by it and enter into the recycling reaction $\mathrm{^{12}C(n,\gamma)
  ^{13}C(\alpha,n)^{16}O}$.   At  the   beginning  of  the  proton  ingestion,
$\mathrm{^{13}C}$ is  much less abundant than  $\mathrm{^{14}N}$, and neutrons
not captured  by $\mathrm{^{12}C}$ are captured by  $\mathrm{^{14}N}$.  As the
PIE  progresses, the  $\mathrm{^{13}C}$ abundance  increases and  so  does the
$\mathrm{^{13}C/^{14}N}$  abundance  ratio   and  the  neutron  density  (Fig.
\ref{figure:4} and Fig.  \ref{figure:5}). Eventually, $\mathrm{^{13}C/^{14}N}$
abundance  ratio gets  larger than  unity  and more  neutrons are  free to  be
captured  by  heavy elements.   The  $\mathrm{^{13}C/^{14}N}$ abundance  ratio
reaches a maximum value around  $\mathrm{\Delta t\sim 0.18}$ yrs after the PIE
starts.  Hydrogen  abundance is  now large enough  to enhance  the competition
between    $\mathrm{^{13}C(p,\gamma)^{14}N}$   and   $\mathrm{^{13}C(\alpha,n)
  ^{16}O}$.   Thus,  the $\mathrm{^{13}C/^{14}N}$  abundance  ratio starts  to
decrease,   leading  to  a   shallower  increase   in  neutron   density  than
before. After  the splitting, the bottom  of the HeCZ  moves slightly outwards
and  cools  down  and  the  supply  of fresh  protons  into  the  HeCZ  stops,
consequently the neutron flux is quickly suppressed (Fig. \ref{figure:4}).

 \begin{figure}
  \resizebox{\hsize}{!}{\includegraphics{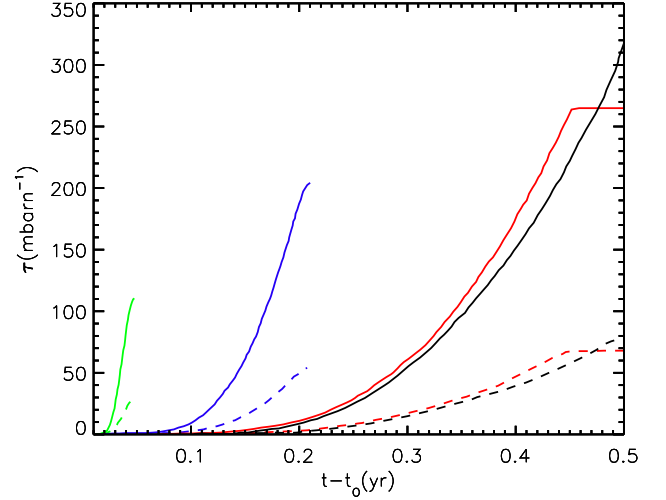}}
  \caption{Time  evolution of  neutron  exposure.  Solid  lines represent  the
    neutron exposure at the position of maximum neutron density (i.e. near the
    bottom  of  the HeCZ)  and  dashed  lines  represent the  average  neutron
    exposure   in    the   convective   zone.      Red   line:   standard
      $\mathrm{Z=10^{-8}}$ 
    model  (M2), black line:  reduced convection  efficiency (M3),  blue line:
    model including overshooting (f=0.016 - M4), and green line: model
      including overshooting (f=0.07 - M5).
  \label{figure:6}}
\end{figure}

Regarding the production  of s-process elements in a  given environment (which
determines, among others, the number of seed nuclei present), the key quantity
is the time-integrated neutron flux, or neutron exposure, given by
\begin{equation}
\tau = \int{v_T n_n dt}
\label{eq:neuexp}
\end{equation}
where  $v_T$ is  the  thermal velocity  of  neutrons, and  $n_n$ their  number
density. In the case of the  PIE, the other relevant quantity is the splitting
timescale $\mathrm{\Delta t_{PIE}}$ which, as we discuss below, determines the
timespan over which  neutrons are effectively produced  and therefore the
  cutoff to the integration in Equation~\ref{eq:neuexp}.

In Figure~\ref{figure:6} we  show, by solid lines,  the neutron exposure
as a function of  time near the bottom of the HeCZ,  where the maximum neutron
flux is  achieved. Results for the  reference model, M2, are  depicted in red.
In the same  figure, dashed lines denote the  neutron exposure averaged across
the entire  convective zone.  For  comparison, typical values for  the neutron
exposure during the AGB phase are roughly two orders of magnitude smaller than
the maximum neutron exposure at the bottom of the HeCZ.

\begin{figure}
  \resizebox{\hsize}{!}{\includegraphics[angle=90]{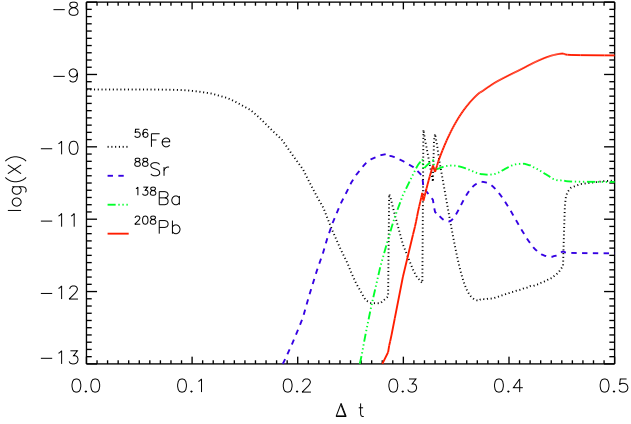}}
  \caption{Evolution of  the mass fraction  of isotopes representative  of the
    three s-process  abundance peaks and  $\mathrm{^{56}Fe}$ as a  function of
    time.    These    abundances   were   sampled   at    the   position   0.4
    $\mathrm{M_{\odot}}$.   They are representative  of the  entire convective
    zone and,  after the splitting, are  located in the HCZ  which later comes
    into contact with the stellar envelope. The peaks, most noticeable for
    $\mathrm{^{56}Fe}$, result  from episodic advance of  the outer convective
    boundary due to limitations in the time and spatial resolution.}
  \label{figure:7}
\end{figure}

At the beginning of the PIE iron-peak elements (mainly $\mathrm{^{56}Fe}$) are
converted  to elements  of the  first  s-process peak  (Sr, Y,  Zr).  The
  $\mathrm{^{56}Fe}$ abundance is strongly reduced, reaching its minimum value
  at about 0.27~yrs after the maximum helium-burning luminosity.  Between 0.27
  and 0.32 yrs after the  He-flash second s-process peak elements are produced
  followed shortly  after by producion of  the heaviest elements  (Pb and Bi).
  The mass fraction of Fe, Sr, Ba,  and Pb in the convective zone are shown in
  Figure~\ref{figure:7}.  The spikes seen in the Fe abundance reflect episodic
  advances  of the  convective shell  over the  radiative upper  layers, which
  dredge-down fresh  material. Near the  end of the PIE,  around $t-t_0\approx
  0.45$~yr, the outer boundary of  the HCZ advances, freshly dredged-down iron
  increases the $\mathrm{^{56}Fe}$ abundance by  an order of magnitude but the
  neutron density simultaneously drops  several orders of magnitude as already
  shown in Fig.~\ref{figure:4}, and the s-process is quenched.

After the  splitting of the HeCZ,  the upper convective  zone, HCZ, penetrates
further into the  H-rich layers above and $\mathrm{^{12}C}$  is converted into
$\mathrm{^{14}N}$.   The   cooling  of   the  HCZ  effectively   switches  off
$\alpha$-captures by $\mathrm{^{13}C}$, and $\mathrm{^{14}N}$ captures most of
the      remaining     neutrons      available,      effectively     quenching
s-processing. Therefore,  the abundances of  heavy elements in the  HCZ remain
unchanged by nuclear burning after the splitting. They do not change until the
HCZ merges with the stellar  envelope and the processed material is dredged-up
to the surface.

\subsection{The influence of convective mixing}
 
The treatment  of convection  and the associated  mixing in  stellar evolution
calculations is subject to fundamental  uncertainties that can not be properly
addressed in 1D models. Nevertheless, we can try to understand how varying the
mixing efficiency  in convective regions or  the extent of  the mixing regions
(e.g. by  including overshooting) affect the  basic properties of  the PIE. In
turn,  this  will  affect  neutron  production and  the  ensuing  creation  of
s-process elements.

Let us  first discuss models including  overshooting. Model M4  is computed by
using a moderate amount of overshooting characterized by $f=0.016$ (Sect.~\ref
{sec:models}). The extended convective boundary  in this model allows a larger
penetration  of  the HeCZ  into  H-rich  layers.  In this  model,  the
  splitting of the HeCZ occurs at $\mathrm{M^H_{max}=0.244~M_\odot}$, somewhat
  deeper  than  for model  M2  for which  $\mathrm{M^H_{max}=0.259~M_\odot}$.
More fuel is ingested into the  HeCZ and thus a faster, but shorter, evolution
of the  proton ingestion episode is  achieved. Comparison of models  M2 and M4
(red and blue lines respectively  in Figure~\ref{figure:6}) show that the more
vigorous entrainment of  hydrogen into the HeCZ in model  M4 produces an early
and steeper increase in the  neutron exposure.  Also, the maximum neutron flux
is almost a factor  of two larger in model M4 than  in model M2.  However, the
faster release of nuclear energy by  hydrogen burning leads to the PIE lasting
less than half the time it does  in the case of no overshooting. This leads to
a significantly  smaller final neutron exposure in  the case of M4  as seen in
Figure~\ref{figure:6}. Model  M5, with a  large overshooting parameter
  $f= 0.07$, confirms this trend. For this model, results 
shown  in green in  Figure~\ref{figure:6} confirm  that a  larger overshooting
region leads  to a more violent hydrogen  ingestion (as seen by  the quick and
steep  rise in  neutron  production)  but of  shorter  duration and,  overall,
reduced  efficiency  in  the  final  neutron  exposure.   Quantitatively,  the
difference  between  the  final  neutron  exposure for  the  two  models  with
overshooting  is  not  too  large,  despite  the  $\sim  50\%$  difference  in
$\mathrm{\Delta  t_{PIE}}$   (see  blue  and  green  solid   lines  in  Figure
\ref{figure:6}).
    
As mentioned  before, we  have found shorter  splitting timescales  than other
recent calculations  (e.g. those of  \citealt{Campbell10}). In order  to mimic
these  longer  splitting  timescales,  we  computed a  model  with  a  reduced
convective mixing efficiency. In our calculations we have simply achieved this
by dividing the diffusion coefficient by  a constant factor.  In model M3 this
reduction is by one order of magnitude. The slower mixing is reflected
  by the outer location of the splitting, at =$\mathrm{M^H_{max}=0.303~M_\odot}$.
As  expected,  a reduced  mixing
efficiency  results  in  a  larger  $\mathrm{\Delta  t_{PIE}}$,  although  the
difference is not too significant compared to the standard case (see Models M2
and M3 in Table \ref{table:1}). This increase in time
produces  a total neutron  exposure at  maximum (black  solid line)  about 20\%
larger than  in the  standard case;  this should favor  the production  of the
heaviest   s-process    isotopes,   as   final   Pb    abundances   given   in
Table~\ref{table:2} show.

In a  radiative environment, the neutron  exposure at the  position of maximum
neutron  density would  result  in a  strong  enhancement in  Pb,  due to  the
repeated neutron captures in a static environment, in all cases. However, in a
convective  environment,  the  interplay  between burning  and  convection  is
important. The mixing timescale is shorter than the overall phase of s-process
production and,  thus, the relevant quantity  for the final  production is the
averaged neutron  exposure. Figure  \ref{figure:6} shows the  neutron exposure
averaged over the convectively mixed zone  (dashed lines).  As it can be seen,
the averaged  neutron exposure is  strongly reduced in the  overshooting model
(blue dashed line), but it is not much affected by the reduction in the mixing
efficiency (black dashed  line). This leads to small  differences in the final
production of  s-process elements despite  the 50\% difference in  the neutron
exposure at the bottom of the HeCZ.

\subsection{The dependence on metallicity}

{During the PIE, the amount of available neutrons do not depend, at least
directly,  on  the initial  metallicity  of  the  star because  the  necessary
$\mathrm{^{12}C}$  is  produced  by  the  helium-core  flash.  Similarly,  the
abudance   of   the   most   relevant   neutron  poison   in   this   context,
$\mathrm{^{14}N}$,  does  not  depend  on  the  initial  metallicity.  On  the
contrary, the number of seed nuclei is a linear function of metallicity
(assuming the  distribution of relative  abundances is not  altered). However,
the development of the PIE may be  altered by the metallicity of the model. We
have found for model Z7, with $Z=10^{-7}$, a slightly more gentle PIE event in
which the  $\mathrm{\log{L^{H}_{max}/L_{\odot}}} = 10.37$, more  than a factor
of two lower than for model M2,  from which it differs only in the initial $Z$
value adopted.
However, $\mathrm{\Delta t_{PIE}=2.2 
  \mbox{yr}}$, a factor of 5 longer and the integrated neutron exposure at the
point   of  maximum   neutron  density   reaches   a  very   large  value   of
370~$\mbox{mb}^{-1}$    and    the    averaged    total    neutron    exposure
100~$\mbox{mb}^{-1}$. These values are somewhat  larger than for model M2. The
production of s-process elements reflects  the scenario depicted above, and it
is roughly one  order of magnitude larger as a result  of the larger abundance
of seed  nuclei but almost  the same of  poison elements in comparison  to the
more metal poor model M2. We come back to this in Sect.~\ref{sec:discussion}.

\subsection{The absence of iron-peak seeds: zero metallicity case}

\citet{Goriely01}  have studied  the occurrence  of  s-process nucleosynthesis
during the  AGB phase  of a $\mathrm{3M_{\odot}}$  stellar model  with initial
zero  metallicity.  They  found that,  by parametrizing  the formation  of the
$\mathrm{^{13}C}$-pocket during the third  dredge-up, neutron densities of the
order of $\mathrm{10^{9}cm^{-3}}$ can  be reached.  Furthermore, their results
showed this is  high enough for an efficient  production of s-process elements
to occur starting  from lighter seeds (C-Ne), even in  the complete absence of
iron-peak elements.

The calculations  of the $\mathrm{1~M_{\odot}}$ metal-free  model presented in
this  paper  (model M1),  show  that low-mass  stars  can  also produce  light
elements that might act as  seeds of s-process nucleosynthesis. This occurs in
two steps.   First, during the transition  from the subgiant to  the red giant
phase, the  interior temperature  is high enough  to start  helium-burning and
produce   carbon.   Once  the   carbon  mass-fraction   is  $\mathrm{X_{c}\sim
  10^{-11}}$, the CNO cycle is ignited  and a small thermal runaway occurs due
to   the   sudden   increase   in   the  efficiency   of   the   CNO   burning
\citep{weiss00}. Later, in the PIE, neutrons are captured by $\mathrm{^{12}C}$
and $\mathrm{^{16}O}$ creating Ne and Na isotopes.

In  model  M1, however,  the  HeCZ splits  before  s-process  elements can  be
produced by neutron  capture onto the light seeds and,  since the splitting of
the HeCZ  is followed by  an abrupt decrease  of the availability  of neutrons
(Figure~\ref{figure:4}, top panel), s-process  elements are not formed despite
of the  high neutron  density ($n_n>10^{13} cm^{-3}$)  reached.  In  fact, the
neutron  exposure  achieved at  the  position  of  maximum energy  release  by
He-burning  is $\mathrm{\sim  25~mb^{-1}}$,  which would  be sufficient  to
produce  s-process elements  if  the  neutrons were  captured  in a  radiative
environment,  as  it   happens  during  the  third  dredge-up   in  AGB  stars
\citep{Goriely01}.

\section{Discussion}
\label{sec:discussion}

The results  presented in  the previous sections  show that the  production of
s-process  elements in the  models is  mainly determined  by the  time spanned
between  the  start of  the  PIE  and the  splitting  of  the convective  zone
($\mathrm{\Delta t_{PIE}}$) and the time  evolution of the neutron flux. It is
in this context that we discuss, in this section, our results and compare them
to those from other authors.

\citet{Fujimoto00} proposed a  general picture for the enrichment  of CNO and,
potentially, s-process  elements in  extremely metal-poor stars.   Although in
their  models  stars   with  mass  $\mathrm{M<1.2M_{\odot}}$  and  metallicity
$\mathrm{[Fe/H]<-4}$  experience  the  PIE   during  the  core  He-flash,  the
splitting  timescale for  the HeCZ  is  so short  that there  is no  s-process
enhancement in  the surface. In  their models related  to the PIE in  the core
He-flash,  based  on those  by  \citet{Hollowell},  they find  $\mathrm{\Delta
  t_{PIE}}\approx2.5\times 10^{-3}  \hbox{yr}$. In more recent  results by the
same group \citep{Suda10}, the picture  remains similar to the one proposed by
\citet{Fujimoto00},   although  there  is   no  available   information  about
$\mathrm{\Delta t_{PIE}}$. On the other hand, \citet{Campbell10} found a large
s-process  production  during  the  PIE  of  a  $\mathrm{M=1~M_{\odot}}$  with
metallicity  $\mathrm{[Fe/H]=-6.5}$. Our  models for  the core  He-flash agree
with  results by  \cite{Campbell10} in  terms  of the  existence of  s-process
production during  this evolutionary  stage.  The level  of production  and of
surface enrichment are, however, quantitatively different.

It  should  be stated  that  an  important  difference among  calculations  by
different  authors relates  to the  treatment of  mixing during  the  PIE.  As
described before, our  calculations with GARSTEC use a  diffusive approach for
convective mixing  to account for  the competition between mixing  and nuclear
burning.  This  is similar to the  scheme used by  \citet{Campbell10}.  On the
other hand,  models computed by \citet{Fujimoto00}  and \citet{Suda10} assume,
during  the PIE,  homogeneous (instant)  mixing down  to a  depth in  the HeCZ
determine  by  equating  the  convective  turn-over  and  proton  captures  by
$\mathrm{^{12}C}$ timescales. Is it this less physical approach at the root of
the   qualitative    differences   in   the    results   regarding   s-process
nucleosynthesis?

We do not find in our models the neutron superburst present in
  \citet{Campbell10} calculations  in which, at beginning of  the PIE, neutron
  density values of $ \mathrm{\sim 10^{15} cm^{-3}}$ are reached, leading 
to the immediate production of large amounts of s-process elements. In our
models, as  the ingestion of  protons progresses, the  neutron density
increases smoothly  up to values above $\mathrm{10^{14}  cm^{-3}}$ (see Figure
\ref{figure:4}).  This results in a  slower production of s-process in the HeCZ
than  the  one found  by  \cite{Campbell10}.  Also,  the  larger  size of  our
convective  zone  contributes  to  effectively  slow down  the  production  of
s-process  elements.  The  combination of  a slower  production and  a shorter
timescale  for the  splitting leads  to  surface abundances  lower than  those
reported by  \citet{Campbell10}: for instance,  their Sr surface  abundance is
more than  2 dex larger  than ours (Table~\ref{table:2}).  However, a
  direct  comparison  with our  results  is  difficult  because production  of
  s-process elements depend also on average properties of the neutron exposure
  in   the   convective   shell,   and   no   information   is   provided   by
  \citet{Campbell10}. Also, we note  here that
\citet{Campbell10}  had  to extrapolate  their  results  because of  numerical
problems  which  did not  allow  them to  follow  the  dredge-up of  processed
material into the stellar envelope.  On the contrary, our results are based on
full models that  follow in detail the enrichment of the  stellar surface as a
consequence  of  the merging  between  the  HCZ  and the  convective  envelope
following the  PIE.  Although we  do not believe  the difference in  the final
enrichement are  caused because  of extrapolations done  by \citet{Campbell10}
(which  is based  on reasonable  assumptions),  it is  clear our  calculations
represent  a qualitative  step forward  in  modelling the  PIE and  subsequent
enrichment of the stellar envelopes.

 We have compared our models with the hyper metal-poor (HMP) stars HE0107-5240
 and  HE1327-2326.  The first  conclusion is  that, in  agreement with
   previous 
 works,  the self-pollution  scenario  cannot explain  the abundance  patterns
 observed  in  HE0107-5240 \citep{Picardi,Suda04,Weiss04,Campbell10}.
 Models produce $\sim 3$  dex more [C/Fe] and at least 2  dex more [Sr/Fe] and
 [Ba/Fe] than  observed in  HE0107-5240.  This  is shown in  the top  panel of
 Fig.~\ref{figure:8}, where abundances derived  from observations are shown in
 blue  circles.   Surface  abundances  of  model  M2,  after  dredge  up,  are
 represented by  the black solid line.   Note that model  abundances have been
 shifted to  be on the scale  determined by [Fe/H]=-5.3  corresponding to this
 star, about  one order  of magnitude larger  than model M2.   Additionally we
 have computed model Z7, with  a solar scaled metallicity $Z=10^{-7}$, closely
 matching [Fe/H]  value for  HE0107-5240.  Results are  shown in  dashed black
 line, and  lead to similar conclusions  as for model  M2.  The self-pollution
 scenario is  not even applicable  to HE1327-2326 because  this star is  not a
 giant and has  not undergone the He-core flash yet.   In the binary scenario,
 on the other  hand, there is no restriction to the  evolutionary stage of the
 observed star and we can compare our models to both stars.  In this scenario,
 material  is transferred  from  the  primary star  (represented  here by  our
 models) and  diluted in the envelope  of the companion  (observed) star.  The
 dilution factor,  $\mathrm{f_D}$, represents the fractional  mass of accreted
 material  with respect  to the  mass  of the  accreting star  over which  the
 polluting material is mixed. Simply,
\begin{equation}
\epsilon^*_i= \frac{\epsilon^{\rm ini}_i+f_D \, \epsilon^{\rm PIE}_i}{1+f_D}, 
\end{equation}
where, for a reference element $i$, $\epsilon^*_i$ is the observed abundance, 
 $\epsilon^{\rm  ini}_i$  the  initial  model  abundance,  and  $\epsilon^{\rm
  PIE}_i$ the abundance after the dredge up that follows the occurrence of the
PIE\footnote{$\mathrm{\log{\epsilon_i} = \log{N_i/N_H}+12}$.}. 
Here, we use  carbon as the reference element to compute  $f_D$ and assume the
initial carbon abundance is the same in both binary components.

Let us first consider HE0107-5240.  For models M2 and Z7 shown in top panel of
Fig.~\ref{figure:8}, the dilution factors are $\mathrm{f_D=7.4\times 10^{-3}}$
and  $\mathrm{f_D=5.4\times  10^{-3}}$  respectively. The  abundance  patterns
after applying  dilution are represented by  blue solid (M2)  and dashed lines
(Z7). The vertical displacement between the models for the Fe abundance simply
reflects  the  different initial  metallicities  of  the  models.  The  almost
constant difference between  models M2 and Z7 for all  elements higher than Al
reflect  that production  of those  elements scales  almost linearly  with the
initial  composition  of  the  models.   Models M2  and  Z7,  after  including
dilution, produce oxygen in agreement with the observed values within a factor
of two, while  N is overproduced by a  factor of $\sim 27$ (M2)  and $\sim 12$
(Z7). For  elements observed  in this star,  in the  range between Mg  and Ni,
there is a  very modest overproduction of elements in  our models with respect
to  Fe.  After  dilution, the  resulting abundance  pattern  for these
elements shown in  Fig.~\ref{figure:8} is almost flat and  reflects mostly the
initial  solar  abundance  pattern   of  the  models.   The  dilution  factor,
determined by the  extremely large carbon enhancement, is too  small to have a
large impact on  these elements. For the s-process elements Sr  and Ba, and Eu
as  well,  for  which  observations   only  allow  an  upper  limit  abundance
determination, the situation is  similar. Although these elements are strongly
overproduced  during  the PIE  (see  Table~\ref{table:2}  and  black lines  in
Fig.~\ref{figure:8}), abundances after
dilution show a pattern very close to solar, i.e. reflect the initial relative
distribution of abundances in the models.   In this regard, if we assume these
elements  were not present  in the  initial composition  of the  star, results
shown in Fig.~\ref{figure:8} can be taken  as an upper limit of the abundances
that the models presented in this work and the pollution scenario can
lead to. 

In the case of  HE1327-2326 we adopt $\mathrm{[Fe/H]}=-5.96$ \citep{Frebel08},
which  is quite close  the Fe  abundance in  our reference  model M2.   In the
bottom  panel of  Fig.~\ref{figure:8} results  for model  M2 before  and after
dilution  ($\mathrm{f_D=1.3\times  10^{-3}}$)  are  compared to  the  observed
abundances. In addition,  to show the impact of  uncertainties in the modeling
of the  PIE, we also  show results for M5  ($\mathrm{f_D=2.1\times 10^{-3}}$),
the  model with  very large  overshooting. There
is a reasonably  good agreement for N  and O abundances, within a  factor of 2
for N and 4 for O.  For all other elements, overproduction factors are totally
damped by the  smallness of the dilution factors,  except for $\mathrm{Ba}$ in
model M5.  

One  interesting signature  of  the PIE  is  the $\mathrm{^{7}Li}$  production
\citep{Iwamoto,Cristallo09}.  Unfortunately, observations  only allow an upper
limit determination  of its abundance in  these two stars.  For HE1327-2326 in
particular,  our  models lead  to  a  value that  agrees  very  well with  the
observational  upper limit.  For  HE0107-5240,  on the  other  hand, model  Z7
overproduces $\mathrm{Li}$  by more than one order  of magnitude.  Our models
do  not account  for any  possible destruction  of $\mathrm{Li}$  after mass
transfer.}

\begin{figure}
  \resizebox{\hsize}{!}{\includegraphics{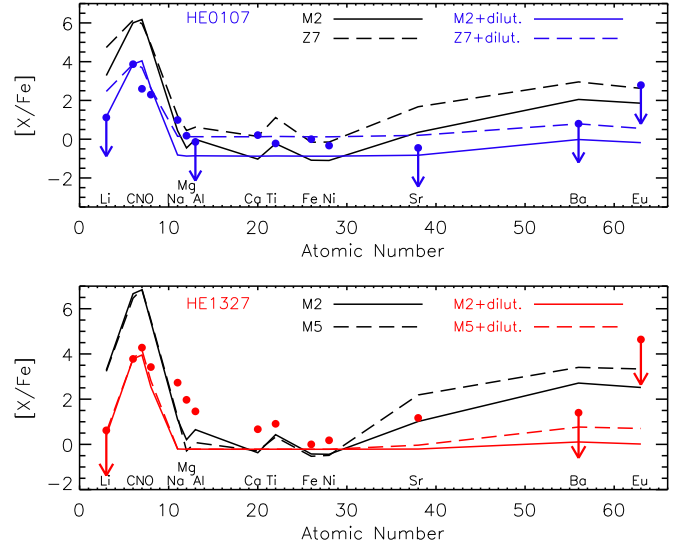}}
  \caption{Top  panel: abundance  pattern  of  HE0107-5240 (blue  filled
      circles, from \citealt{Chris04}  and \citealt{Bessel}).  Black solid and
      dashed  lines represent  surface  abundance  of models  M2  and Z7  (see
      Sect.~\ref{sec:discussion} for more details on this model) respectively.
      Blue  lines show the abundance patterns  after dilution  by factors of
      $\mathrm{f_{D}=7.4\times 10^{-3}}$  and $5.4 \times  10^{-3}$ (models M2
      and Z7 respectively) derived  by matching the observed carbon abundance.
      Arrows   indicate   upper   limits.     Li   abundance   is   given   as
      $\mathrm{log\epsilon(Li)=log(N_{Li}/N_{H})+12}$.  Zero point on
        the       y-axis corresponds to [Fe/H]=-5.3}. Bottom panel: same as
      above   but   for   HE-1327-2326   (red  filled   circles,   data   from
      \citealt{Frebel08}).   For  this  star,  models  shown are  M2  and  M5.
      Abundance  patterns after  dilution  ($\mathrm{f_{D}=1.3\times 10^{-3}}$
      and $2.1\times  10^{-3}$ for  M2 and M5  respectively) are shown  in red
      solid and dashed lines as  indicated in the plot. Zero point of
        the       y-axis corresponds to [Fe/H]= -5.96. \label{figure:8}
\end{figure}

In our simple dilution model, we have assumed that mass was transferred before
the star entered  the AGB phase.  Low-mass stars that  suffered PIE during the
core He-flash  undergo a normal  AGB evolution, i.e.,  there is no  PIE during
this phase. Therefore, the production of s-process elements should happen in a
radiative  environment  if a  $\mathrm{^{13}C}$  pocket  is  formed. We  might
speculate on the possible consequence  if the primary had not transferred mass
to the companion star  right after the PIE but rather during  or after its AGB
evolution. The main products carried  to the surface after the third dredge-up
(TDU) are carbon and s-process  elements. Hence, before transferring mass, the
primary  would be  even more  enriched in  these elements  than prior  to this
phase. Nitrogen abundance would remain the same since Hot Bottom Burning (HBB)
does not happen in stars  with masses $\mathrm{M=1~M_{\odot}}$ and Li would be
depleted during the TDU.  In this way,  a better match to the nitrogen and the
lithium abundances of  HE 0107-5240 could be obtained if  the accreted mass is
small enough.  Also, the  $\mathrm{^{12}C/^{13}C}$ ratio would increase during
the  AGB evolution  and could  better match  the observed  lower limit  for HE
0107-5240 ($\mathrm {^{12}C/^{13}C>50}$).

\begin{table}
\caption{Surface abundances}    
\label{table:2}     
\centering                               
\begin{tabular}{cccccc}       
\hline\hline                       
Element\tablefootmark{a} & M2 & M3 & M4 & M5 & Z7\\
\hline
    C & 6.97 & 6.80 & 6.99 & 6.90 & 6.25\\
    N & 7.05 & 6.73 & 7.04 & 7.20 & 5.91\\
    O & 5.64 & 5.46 & 5.69 & 5.81 & 4.83\\
    Sr & 1.49 & 1.59 & 1.86 & 2.78 & 1.88\\
    Y & 1.86 & 1.91 & 2.04 & 2.59 & 2.23\\
    Zr & 1.92 & 2.05 & 2.21 & 3.17 & 2.33\\
    Ba & 3.16 & 2.77 & 3.06 & 3.96 & 3.13\\
    La & 3.27 & 2.76 & 3.37 & 4.29 & 3.24\\
    Eu & 2.97 & 2.32 & 3.04 & 3.89 & 2.81\\
    Pb & 4.74 & 4.86 & 5.00 & 4.52 & 5.31\\
\hline                                            
\end{tabular}
\tablefoot{
\tablefoottext{a}{All abundances are given in terms of:
  [X/Fe] = $\mathrm{log_{10}(X/Fe)_{\star} - log_{10}(X/Fe)_{\odot}}$.}
}
\end{table}

An important finding of this work is that in the Z=0 model only elements
lighter than  Fe are  synthesized.  Our results  show that the  splitting time
$\Delta {\rm  t_{PIE}}$ of the HeCZ is  too short and not  enough neutrons are
produced to go  beyond Fe. In the viewpoint of  the binary pollution scenario,
this result does not support a low-mass zero metallicity star as the polluting
companion of the two HMP stars discussed in this paper, if mass is transferred
before it enters the AGB evolution.

In closing  the discussion,  we have done  a simple  test to determine  if the
production of  s-process elements during the  PIE is sensitive  to the stellar
mass.  As  stated before,  \citet{Fujimoto00}  has  found  an upper  limit  of
1.2~M$_\odot$ for  the occurrence of the  PIE. On the low-mass  range, we have
computed   the   evolution  of   a   0.82~M$_\odot$   and  Z=10$^{-8}$   model
(representative of the lowest stellar mass  that has had time to evolve to the
core He-flash in the lifetime of  the Milky Way) and found that, in comparison
to   model   M2,   the   splitting   time   is   a   factor   of   5   shorter
($\Delta_{PIE}=0.08~\mathrm{yr}$), but  still more  than 30 times  larger than
models by \citet{Hollowell}.  The neutron flux increases more  rapidly than in
model  M2  and  just  $0.03~\mathrm{yr}$  after  $\mathrm{L^{He}_{max}}$,  the
neutron  density   at  the  bottom  of   the  HeCZ  is   already  larger  than
$\mathrm{10^{14}~cm^{-3}}$. As a result we find that, associated with the PIE,
there   is   a  total   neutron   exposure  at   the   bottom   the  HeCZ   of
$130~\mathrm{mb^{-1}}$ and an average  neutron exposure over the whole HeCZ
of   $30~\mathrm{mb^{-1}}$.   We   have   not   performed   post-processing
nucleosynthesis for this model. However, the abundance pattern during the HeCZ
is very  similar to model  M2 (and common  to all models of  equal metallicity
undergoing the core He-flash) and the neutron exposure is very large and
within a factor of 2 of model M2. For these reasons, and also based on results
presented in this work we conclude that s-process elements will be efficiently
produced  with   a  qualitatively  similar  abundance   pattern  in  extremely
metal-poor stars  in the low-mass  range of stars  suffering a PIE  during the
core He-flash, that is for stellar masses smaller than about $1.2~M_\odot$.

\section{Conclusions}
 
We  have  performed  evolutionary  calculations  for  low-mass  EMP  and  zero
metallicity stars. Our models follow the proton ingestion episode (PIE) during
the core He-flash  and the subsequent hydrogen flash. We  have then used these
calculations  as input  to a  post-processing  unit, where  we have  performed
calculations of s-process nucleosynthesis. A comparison with and among similar
calculations  found in the  literature shows  that the  location at  which the
He-flash   occurs  varies   among  different   stellar  codes   (varying  from
0.15~M$_\odot$  \citep{Schlattl01},   $\sim  0.24$~M$_\odot$  (this paper)  to
0.41~M$_\odot$ \citep{Hollowell}).  This is indicative of the uncertainties in
stellar modeling and is likely to impact results of the PIE modeling.

The production of s-process  elements and subsequent surface enrichment during
the  PIE in  EMP low-mass  stars  depends strongly  on the  efficiency of  the
convective  mixing  and on  the  general  properties  of the  PIE,  in
  particular on  the $\mathrm{\Delta t_{PIE}}$,  the time elapsed  between the
  moment of maximum He-burning luminosity  and the splitting of the convective
  zone.  We do find large production of s-process elements
in our  models. The  surface abundances after  dredge- up are,  however, about
2~dex smaller than those reported by \citet{Campbell10}.  This is likely the
result of the difference in the neutron density history between our models and
those by \citet{Campbell10}; we do not find the neutron superburst reported by
these  authors. A  detailed comparison  with \citet{Campbell10}  is difficult.
Although they report a  total neutron exposure of 287~$\mathrm{mb^{-1}}$, close
  to our  results for models M2  and M3, the  PIE takes place in  a convective
  region and neutron production is a  strong function of temperature, as it is
  determined  by both  the producion  and burning  of  $\mathrm{^{13}C}$.  The
  actual production of  s-process elements, therefore, does not  depend on the
  maximum neutron exposure, but on  its averaged value. No information on this
  regard  is provided by  \citet{Campbell10}, and  neither how  the convective
  region has been treated in their post-processing code.
In  more detail,  our models  produce 2-3~dex  less  first and
second peak s-process elements whereas  Pb production is smaller in our models
by 1.3~dex. On the other  hand, our results qualitatively agree that s-process
production happens  during the  PIE in  the core-He flash  phase, which  is in
contrast  with   \citep{Fujimoto00,Suda10}  results  who   found  no  relevant
s-process production during the PIE.

Our models  produce C,  N, and O  surface abundances  2-3 dex larger  than the
abundance values  observed in the  HMP stars. The  high CNO abundances  in our
models  disfavor the  self-enrichment scenario  to explain  the  EMP abundance
patterns, as other works on the  topic have shown before.  In the viewpoint of
the mass-transfer scenario,  our models support the idea  that a low-mass star
can be the donor star because  the dilution of the transferred material occurs
in  the secondary  star  envelope. In  fact,  our estimation  of the  dilution
factors during the accretion process shows that the secondary needs to accrete
a mass  equal to only  a few parts  per thousand of  its own envelope  mass in
order  to match  the  observed  carbon abundances.    The small  dilution
  factors,  on the other  hand, makes  it difficult  to enhance  the s-process
  abundances of the  accreting star for the overproduction  levels we find and
  report in Table~\ref{table:2}.

Based  on  our   results,  and  in  qualitative  agreement   with  those  from
\citet{Campbell10},  we  conclude that  stars  with masses  $\mathrm{M\leq1~M_
  {\odot}}$ cannot be  excluded as the binary companions of  the two most iron
deficient  stars   yet  observed.   In   addition,  the  latest   findings  by
\citet{Caffau2011,Caffau2012} push  the lowest  metallicity at which  low mass
stars  can form,  increasing the  need of  detailed models  for this  class of
stars.  All these  results reinforce the necessity of  detailed studies of the
PIE, both  with hydrodynamic simulations  and stellar evolutionary  models, in
order to achieve a more realistic picture of the its properties.

 \begin{acknowledgements}
This work  is part  of the  Ph.D. thesis of  Monique Alves  Cruz and  has been
funded by the IMPRS fellowship. This research was supported by the DFG cluster
of  excellence ``Origin  and Structure  of the  Universe''. Aldo  Serenelli is
partially supported  by the  European Union International  Reintegration Grant
PIRG-GA-2009-247732 and the MICINN grant AYA2011-24704.  M.A.C.
would like to thank N.M.  Fernandes  for the careful reading of the manuscript
and Silvia Rossi for providing her with a nice work space for the past months.
We acknowledge the numerous comments  by an anonymous referee that have helped
us improved presentation of results.
 \end{acknowledgements}

\bibliographystyle{aa} 
\bibliography{msrv3} 

\end{document}